\documentclass[12pt]{article}
\usepackage{ctex}
\usepackage{amssymb,amsmath,color,mathdots}
\usepackage[colorlinks,linkcolor=blue]{hyperref}
\usepackage{geometry}
\usepackage{tabularx}
\usepackage{bm}
\usepackage{authblk}
\usepackage{graphicx}
\usepackage[square, comma, sort&compress, numbers]{natbib}

\title{\bf Several classes of projective few-weight linear codes and their applications}	
\author{\small Canze Zhu}
\author{\small Qunying Liao
	\thanks{Corresponding author.
		
		{~E-mail. qunyingliao@sicnu.edu.cn (Q. Liao), ~canzezhu@163.com (C. Zhu).}	
		
		{~Supported by National Natural Science Foundation of China (Grant No. 12071321).}}
}
\affil[]{\small (College of Mathematical Science, Sichuan Normal University, Chengdu Sichuan, 610066)}
\date{}
\newtheorem{theorem}{Theorem}[section]

\newtheorem{lemma}{Lemma}[section]

\catcode`@=11 \@addtoreset{equation}{section} \catcode`@=12
\begin{document}
	\maketitle
		{\bf Abstract.}
	{\small In this paper, let $q$ be a power of the prime, we construct several classes of projective few-weight linear codes over  $\mathbb{F}_q$ from  defining sets, and determine their  weight distributions by using character sums. Especially, we obtain three classes of Griesmer codes or distance-optimal near Griesmer codes, and two classes of projective two-weight codes. Furthermore, two  classes of strongly regular graphs are given from  these projective two-weight linear codes and some of these codes are suitable for applications in secret sharing schemes.}\\
	
	
	\section{Introduction}
	Let $\mathbb{F}_{q^m}$ be the finite field with $q^m$ elements, where $q$ is a power of the prime $p$ and $m$ is a positive integer. An $[n,k,d]$ linear code $\mathcal{C}$ over $\mathbb{F}_q$ is a $k$-dimensional subspace of $\mathbb{F}_q^n$ with minimum (Hamming) distance $d$ and length $n$. 
	The dual code of $\mathcal{C}$ is defined as 
	$$	 \mathcal{C}^{\perp}=\{\mathbf{c}^{'}\in\mathbb{F}_q^n~|~\langle\mathbf{c}^{'},\mathbf{c}\rangle=0~ \text{for any}~\mathbf{c}\in\mathcal{C}\}.$$
	Clearly, the dimension of $\mathcal{C}^{\perp}$ is $n-k$. a linear code $\mathcal{C}$ is said to be projective if the minimum distance of $\mathcal{C}^{\perp}$  is greater than or equal to $3$.  $\mathcal{C}$ is said to be distance-optimal
	if there is no an $[n,k,d+1]$ linear code over $\mathbb{F}_{q}$. $\mathcal{C}$ is called optimal if its parameters $n$, $k$ and $d$ meet one of the bound such as the Plotkin
	bound, the Singleton bound or the Griesmer bound  \cite{WC2003}. Especially, the Griesmer bound is given by
	\begin{align}\label{GB}
	n\ge\sum_{i=0}^{k-1}\lceil\frac{d}{q^{i}}\rceil,
	\end{align}
	where $\lceil\cdot\rceil$ denotes the ceiling function. $\mathcal{C}$ is called a Griesmer code (resp. near Griesmer code) if its parameters $n$ (resp. $n−1$), $k$ and $d$ achieve the Griesmer bound.  
	Griesmer codes have
	been attracted a lot of attention due to  their optimality and geometric applications \cite{CDB2015,CD2018}.
	Normally it is a hard problem to construct optimal linear codes, the reader can refer \cite{H,R1,R2,R3,R4} for recent results.
	
	Let $A_i (i=1,\ldots,n)$ be the number of codewords with Hamming weight $i$ in $\mathcal{C}$, then the weight  enumerator  of $\mathcal{C}$ is defined by the polynomial
	$1+A_1z+A_2z^2+\cdots+A_nz^n$ and the sequence $(1,A_1,\ldots,A_n)$  is called its weight distribution. $\mathcal{C}$ is called a $t$-weight code if the number of nonzero $A_i$ in the sequence $(A_1,A_2,\ldots,A_n)$ is equal to $t$. The weight distribution is an important parameter for a linear code, which can be applied to determine the  capability for both error-detection and error-correction \cite{KT2007}.  In addition, few-weight linear codes have been better applications in secret sharing schemes \cite{JY2006,CC2005},  association schemes \cite{AC1984}, authentication codes \cite{CD2005}, and so on.  In particular, projective two-weight codes are very precious as they are closely related to finite projective spaces, strongly regular graphs and combinatorial
	designs \cite{RC1986,CD2018,P1972}. However, projective two-weight codes are rare and only a few classes are known \cite{RC1986,ZH20161,ZH2017,ZH2021,CD2007,CD2008,CD2015}.
	
	In 2015, Ding et al. constructed a class of two-weight or three-weight linear codes via the trace function from defining sets \cite{KD2015}. Let 
	$D=\{ d_1, d_2,\ldots, d_n \}\subseteq\mathbb{F}_{q^{m}}^*$ and $\mathrm{Tr}^{q^{m}}_{q}(\cdot)$ denote the trace function from $\mathbb{F}_{q^{m}}$ to $\mathbb{F}_q$. A linear code over $\mathbb{F}_q$ is defined by
	\begin{align*}
	\mathcal{C}_D=\big\{\mathbf{c}(x)=\big(\mathrm{Tr}^{q^{m}}_{q}(xd_1),\mathrm{Tr}^{q^{m}}_{q}(xd_2),\cdots,\mathrm{Tr}^{q^{m}}_{q}(xd_n)\big)~\big{|}~ x\in\mathbb{F}_{q^{m}}\big\}.
	\end{align*} 
	Motivated by the above construction, for any nonempty sets $S\subseteq\mathbb{F}_{q}^{m_1}$ and $D\subseteq\mathbb{F}_{q}^{m_2}$, the bivariate form of the defining set construction is given by
	
	\begin{align}\label{CD}
	\mathcal{C}_{S\times D}=\big\{\big(\mathrm{Tr}^{q^{m_1}}_q(ax)+\mathrm{Tr}^{q^{m_2}}_q(by)\big)_{(x,y)\in \mathrm{S}\times D}~\big{|}~(a,b)\in\mathrm{F}_{q^{m_1}}\times\mathrm{F}_{q^{m_2}}\big\}.
	\end{align}	
	The defining set construction has been attracted much attention and there are a lot of few-weight linear codes have been constructed \cite{KD2014, KD2015, SY2015, ZH2015, ZH2016, ZH20161, ZH20162, GJ2019, CL2016, GL2018, CT2016, ZZ2015, Z, CS2019,ZH2017,ZH2021,CD2007,CD2008,CD2015}.
	
	In this paper, let \begin{align}\label{S}
	S=\Big\{\alpha^{i}~\Big|~i=1,\ldots,\frac{q^{m_1}-1}{q-1}\Big\},
	\end{align} where $\alpha$ is a primitive element in $\mathbb{F}_{q^{m_1}}$. By choosing some special set 
	\begin{align}\label{D}
		D\subseteq\mathbb{F}_{q^{m_2}}^{*} \text{~with~} zD=D~ (\forall z\in\mathbb{F}_q^{*}),
	\end{align}  we obtain several classes of projective few-weight linear codes from the construction $(\ref{CD})$, and determine their weight distributions by using character sums. We have the following contributions:
	
	$(1)$ Let $D=\mathbb{F}_{q^{m_2}}^{*}$ or $\mathbb{F}_{q^{m_2}}$, then $\mathcal{C}_{S\times D}$ is not only a projective two-weight or three-weight linear code, but also a Griesmer code or distance-optimal near Griesmer code. Furthermore, two classes of corresponding strongly regular graphs are given.
	
	$(2)$ Let $D=\big\{y\in\mathbb{F}_{q^{m_2}}^{*}|\mathrm{Tr}^{q^{m_2}}_q(y)\neq 0\big\}$ or $\big\{y\in\mathbb{F}_{q^{m_2}}^{*}|\mathrm{Tr}^{q^{m_2}}_q(y)\neq 0\big\}\cup\{0\}$,
	then $\mathcal{C}_{S\times D}$ is a projective four-weight or five-weight linear code. Especially, a class of near Griesmer codes is given.
	
	$(3)$ For odd $q$, let $D=\big\{y\in\mathbb{F}_{q^{m_2}}^{*}|\mathrm{Tr}^{q^{m_2}}_q(y^2)= 0\big\}$ or $\big\{y\in\mathbb{F}_{q^{m_2}}^{*}|\mathrm{Tr}^{q^{m_2}}_q(y^2)= 0\big\}\cup\{0\}$, then
	$\mathcal{C}_{S\times D}$ is a projective $t$-weight $(t\in\{5,6,7\})$ linear code. Especially, when $m_2\ge 8$, $\mathcal{C}_{S\times D}$ is a minimal code. Therefore, its dual code can be employed to construct secret sharing schemes with interesting access structures.  
	

	This paper is organized as follows. In Section 2, some related basic notations and
	results of character sums are given. In Section 3, the weight distributions of several classes of projective few-weight linear codes are presented. In Section 4, the proofs for the main results are given. In Section 5,  we show that some of these codes can be applicated for secret sharing schemes, and some strongly regular graphs with new parameters are derived basing on these projective two-weight codes. In Section 6, we conclude the whole paper.
	\section{Preliminaries}
	
	
	
	\subsection{Group characters and Gauss sums}
	For $s_1$, $s_2$ $\in$ $\mathbb{Z}^{+}$ with $s_1\mid s_2$, $\mathrm{Tr}^{p^{s_2}}_{p^{s_1}}(\cdot)$ is the trace map from $\mathbb{F}_{p^{s_2}}$ to $\mathbb{F}_{p^{s_1}}$, i.e.,  
	\begin{align*}
	\mathrm{Tr}^{p^{s_2}}_{p^{s_1}}(x)=x^{p^{s_2-s_1}}+x^{p^{s_2-2s_1}}+\cdots+x^{p^{s_1}}\quad(\forall x\in\mathbb{F}_{p^{s_1}}).
	\end{align*}
	
	An additive character $\chi$ of $\mathbb{F}_{q^m}$ is a function from $\mathbb{F}_{q^m}$ to the multiplicative group $U=\{u\ |\ |u|=1,\ u\in\mathbb{C}\}$, such that $\chi(x+y)=\chi(x)\chi(y)$ for any $x,\ y\in\ \mathbb{F}_{q^m}$. For each $b\in \mathbb{F}_{q^m}$, the function
	\begin{center}
		$\chi_b(x)=\zeta_p^{\mathrm{Tr}^{q^{m}}_p(bx)}$ \quad $(x\in \mathbb{F}_{q^m})$
	\end{center}
	defines an additive character of $\mathbb{F}_{q^m}$. 
	The orthogonal property for the additive character is given by
	\begin{align*}
	\sum\limits_{x\in \mathbb{F}_{q^m}}\zeta_p^{\mathrm{Tr}^{q^m}_p(bx)}=\begin{cases}
	q^m,& \text{if}~b=0;\\
	0,\quad &\text{otherwise}.
	\end{cases}
	\end{align*}     
	We extend the quadratic character $\eta_m$ of $\mathbb{F}_{q^m}^*$ by letting $\eta_m(0)=0$, then the quadratic Gauss sums $G_m$ over $\mathbb{F}_{q^m}$ is defined as 
	\begin{center}
		$G_m=\sum\limits_{x\in \mathbb{F}_{q^m}}\eta_m(x)\chi(x)$.
	\end{center}
	
	Now, some properties for $\eta_m$ and $G_m$ are given as follows.    	
	\begin{lemma}[\cite{KD2015}, Lemma 7]\label{l21}
		For $x\in\mathbb{F}_q^*$, 
		\begin{align*}
		\eta_m(x)=\begin{cases}
		1,& ~\text{ if~$m$ is even};\\
		\eta_{1}(x),\quad &~\text{ if~$m$ is odd}.
		\end{cases}
		\end{align*}     	
	\end{lemma} 
	\begin{lemma}[\cite{RL97}, Theorem 5.15]\label{l22}
		Let $q=p^s$, then
		\begin{align*}
		G_m=(-1)^{sm-1}\sqrt{-1}^\frac{(p-1)^2sm}{4}q^{\frac{m}{2}}.
		\end{align*}
	\end{lemma}
	
	\begin{lemma}[\cite{RL97}, Theorem 5.33]\label{l28}
		Let $f(x)=a_2x^2+a_1x+a_0\in \mathbb{F}_{q^m}[x]$ with $a_2\neq0$, then
		\begin{align*}
		\sum\limits_{x\in \mathbb{F}_{q^m}}\chi(f(x))=G_m\eta_m(a_2)\zeta_p^{\mathrm{Tr}^{q^{m}}_p(a_0-a_1^2(4a_2)^{-1})}.
		\end{align*}
	\end{lemma}

\section{Main results}	

The following theorem shows that $\mathcal{C}_{S\times D}$ constructed in $(\ref{CD})$ with $S$ given in $(\ref{S})$ is projective.
\begin{theorem}\label{t0}
	Let $d^{\perp}$ be the minimum distance of $\mathcal{C}_{S\times D}^{\perp}$, then $d^{\perp}\ge 3$, i.e., $\mathcal{C}_{S\times D}$ is projective.
\end{theorem}

{\bf Proof}. If $d^{\perp}=1$, then there exists some $(x,y)\in S\times D$ such that $\mathrm{Tr}^{q^{m_1}}_q(ax)+\mathrm{Tr}^{q^{m_2}}_q(by)=0$ $(\forall(a,b)\in\mathrm{F}_{p^{m_1}}\times\mathrm{F}_{p^{m_2}})$, which leads $(x,y)=(0,0)$. Thus $d^{\perp}\neq 1$.

If $d^{\perp}=2$, then there exist $\lambda_i\in\mathbb{F}_q^{*}$ and $(x_i,y_i)\in S\times D$ $(i=1,2)$ with $(x_1,y_1)\neq (x_2,y_2)$ such that $$\lambda_1\Big(\mathrm{Tr}^{q^{m_1}}_q(ax_1)+\mathrm{Tr}^{q^{m_2}}_q(by_1)\Big)+\lambda_2\Big(\mathrm{Tr}^{q^{m_1}}_q(ax_2)+\mathrm{Tr}^{q^{m_2}}_q(by_2)\Big)=0~\big(\forall(a,b)\in\mathrm{F}_{p^{m_1}}\times\mathrm{F}_{p^{m_2}}\big),$$
namely, 
$$\mathrm{Tr}^{q^{m_1}}_q\big(a(x_1+\lambda_1^{-1}\lambda_2x_2)\big)+\mathrm{Tr}^{q^{m_2}}_q\big(b(y_1+\lambda_1^{-1}\lambda_2y_2)\big)=0~\big(\forall(a,b)\in\mathrm{F}_{p^{m_1}}\times\mathrm{F}_{p^{m_2}}\big).$$
Thus $x_1+\lambda_1^{-1}\lambda_2x_2=y_1+\lambda_1^{-1}\lambda_2y_2=0$, which implies $(x_1,y_1)=-\lambda_1^{-1}\lambda_2(x_2,y_2)$. Furthermore, by
$S=\Big\{\alpha^{i}~\Big|~i=1,\ldots,\frac{q^{m_1}-1}{q-1}\Big\}$, we have  $-\lambda_1^{-1}\lambda_2=1$, it leads $(x_1,y_1)=(x_2,y_2)$. Thus $d^{\perp}\neq 2$. 
$\hfill\Box$\\

In the following theorems, we determine the weight distributions for linear codes $\mathcal{C}_{S\times D_i}$ and $\mathcal{C}_{S\times \tilde{D}_i}$ $(i=1,2,3)$, respectively, where $\tilde{D}_i=D_i\cup\{0\}~(i=1,2,3)$,
\begin{align*}
D_1=\mathbb{F}_{q^{m_2}}^{*},~~~~D_2=\big\{y\in\mathbb{F}_{q^{m_2}}^{*}|\mathrm{Tr}^{q^{m_2}}_q(y)\neq 0\big\},~~~~D_3=\big\{y\in\mathbb{F}_{q^{m_2}}^{*}|\mathrm{Tr}^{q^{m_2}}_q(y^2)= 0\big\}.
\end{align*}	

\begin{theorem}\label{t1}
	$\mathcal{C}_{S\times D_1}$ is a ${\Big{[}\frac{(q^{m_1}-1)(q^{m_2}-1)}{q-1}, m_1+m_2,q^{m_1+m_2-1}-q^{m_1-1}-q^{m_2-1}\Big{]}}$ linear code with weight distribution in Table $1$. Furthermore,  if $m_1=m_2$ and $q=2$, $\mathcal{C}_{S\times D_1}$ is a distance-optimal near Griesmer code. Otherwise,  $\mathcal{C}_{S\times D_1}$ is a Griesmer code.
	\begin{center} Table $1$~~~ The weight distribution of  $\mathcal{C}_{S\times D_1}$
		
		\begin{tabular}{|p{6cm}<{\centering}| p{3cm}<{\centering}|}
			\hline   weight $w$	                     &   frequency $A_w$                    \\ 
			\hline       $0$	                     &  $1$                                    \\
			\hline   $q^{m_1+m_2-1}-q^{m_1-1}$   &  $q^{m_1}-1$        \\ 
			\hline   $q^{m_1+m_2-1}-q^{m_2-1}$	                     &      $q^{m_2}-1$  \\
			\hline   $q^{m_1+m_2-1}-q^{m_1-1}-q^{m_2-1}$   &  $(q^{m_1}-1)(q^{m_2}-1)$        \\          
			\hline
		\end{tabular} 
	\end{center}

\end{theorem}

\begin{theorem}\label{t11}
	$\mathcal{C}_{S\times \tilde{D}_1}$ is a ${\Big{[}\frac{(q^{m_1}-1)q^{m_2}}{q-1}, m_1+m_2,q^{m_1+m_2-1}-q^{m_2-1}\Big{]}}$ linear code with weight distribution in Table $2$.
	Furthermore, $\mathcal{C}_{S\times \tilde{D}_1}$ is a Griesmer code.
	
	\begin{center} Table $2$~~~ The weight distribution of  $\mathcal{C}_{S\times\tilde{D}_1}$
		
		\begin{tabular}{|p{6cm}<{\centering}| p{3cm}<{\centering}|}
			\hline   weight $w$	                     &   frequency $A_w$                    \\ 
			\hline       $0$	                     &  $1$                                    \\
			\hline   $q^{m_1+m_2-1}$   &  $q^{m_1}-1$        \\ 
			\hline   $q^{m_1+m_2-1}-q^{m_2-1}$	                     &      $q^{m_1+m_2}-q^{m_1}$  \\      
			\hline
		\end{tabular} 
	\end{center}
\end{theorem}

\begin{theorem}\label{t2}
For $m_2\ge 2$,	$\mathcal{C}_{S\times D_2}$ is a 
${\Big{[}q^{m_1+m_2-1}-q^{m_2-1},m_1+m_2,q^{m_1+m_2-1}-q^{m_1+m_2-2}-q^{m_2-1}\Big{]}}$ linear code with  weight distribution in Table $3$. Furthermore, if $m_2=2$, then $\mathcal{C}_{S\times D_2}$ is a distance-optimal near Griesmer code.
	
	\begin{center} Table $3$~~~ The weight distribution of  $\mathcal{C}_{S\times D_2}$
		
		\begin{tabular}{|p{7cm}<{\centering}| p{4cm}<{\centering}|}
			\hline   weight $w$	                     &   frequency $A_w$                    \\ 
			\hline       $0$	                     &  $1$                                    \\
			\hline   $q^{m_1+m_2-1}-q^{m_1+m_2-2}$   &  $q^{m_1}-1$        \\ 
			\hline   $q^{m_1+m_2-1}-q^{m_1+m_2-2}-q^{m_2-1}+q^{m_2-2}$   &  $q^{m_1+m_2}-q^{m_1+1}$        \\ 
			\hline   $q^{m_1+m_2-1}-q^{m_2-1}$	                     &      $q-1$  \\
			\hline   $q^{m_1+m_2-1}-q^{m_1+m_2-2}-q^{m_2-1}$   &  $q^{m_1+1}-q^{m_1}-q+1$        \\          
			\hline
		\end{tabular} 
	\end{center}
\end{theorem}

\begin{theorem}\label{t21}
	For $m_2\ge 2$, $\mathcal{C}_{S\times \tilde{D}_2}$ is a ${\Big{[}q^{m_1+m_2-1}-q^{m_2-1}+\frac{q^{m_1}-1}{q-1},m_1+m_2\Big{]}}$ 
	 linear code with weight distribution is given in Table $4$.
	\begin{center} Table $4$~~~ The weight distribution of  $\mathcal{C}_{S\times \tilde{D}_2}$
		
		\begin{tabular}{|p{8cm}<{\centering}| p{5cm}<{\centering}|}
			\hline   weight $w$	                     &   frequency $A_w$                    \\ 
			\hline       $0$	                     &  $1$                                    \\
			\hline   $q^{m_1+m_2-1}-q^{m_1+m_2-2}+q^{m_1-1}$   &  $q^{m_1}-1$        \\ 
			\hline   $q^{m_1+m_2-1}-q^{m_1+m_2-2}-q^{m_2-1}+q^{m_2-2}$   &  $q^{m_2}-q$        \\ 
			\hline   $q^{m_1+m_2-1}-q^{m_2-1}$	                     &      $q-1$  \\
			\hline   $q^{m_1+m_2-1}-q^{m_1+m_2-2}-q^{m_2-1}+q^{m_2-2}+q^{m_1-1}$   &  $q^{m_1+m_2}-q^{m_1+1}-q^{m_2}+q$        \\   
			\hline   $q^{m_1+m_2-1}-q^{m_1+m_2-2}-q^{m_2-1}+q^{m_1-1}$   &  $q^{m_1+1}-q^{m_1}-q+1$        \\          
			\hline
		\end{tabular} 
	\end{center}
\end{theorem}

\begin{theorem}\label{t31e}
	If $m_2$ is even, then $\mathcal{C}_{S\times D_3}$ is a ${\Big{[}(q^{m_1}-1)(\frac{q^{m_2-1}-1}{q-1}+\frac{G_{m_2}}{q}),m_1+m_2\Big{]}}$ 
	linear code with weight distribution in Table $5$.
	\begin{center} Table $5$~~~ The weight distribution of  $\mathcal{C}_{S\times {D}_3}$~$(2\mid m_2)$
		
		\begin{tabular}{|p{8.5cm}<{\centering}| p{6cm}<{\centering}|}
			\hline   weight $w$	                     &   frequency $A_w$                    \\ 
			\hline       $0$	                     &  $1$                                    \\
			\hline   $q^{m_1+m_2-2}-q^{m_1-1}+(q-1)q^{m_1-2}G_{m_2}$   &  $q^{m_1}-1$        \\ 
			\hline   $q^{m_1+m_2-2}-q^{m_2-2}$   &  $q^{m_2-1}-1+(q-1)\frac{G_{m_2}}{q}$        \\ 
			\hline   $q^{m_1+m_2-2}-q^{m_2-2}+(q^{m_1}-1)\frac{G_{m_2}}{q}$	                     &      $(q-1)\big(q^{m_2-1}+\frac{G_{m_2}}{q}\big)$  \\
			\hline   $q^{m_1+m_2-2}-q^{m_2-2}-q^{m_1-1}+(q-1)q^{m_1-2}G_{m_2}$   &  $(q^{m_1}-1)\big(q^{m_2-1}-1+(q-1)\frac{G_{m_2}}{q}\big)$        \\   
			\hline   $q^{m_1+m_2-2}-q^{m_2-2}-q^{m_1-1}+\big((q-1)q^{m_1-1}-1\big)\frac{G_{m_2}}{q}$   &  $(q^{m_1}-1)(q-1)\big(q^{m_2-1}+\frac{G_{m_2}}{q}\big)$        \\          
			\hline
		\end{tabular} 
	\end{center}
\end{theorem}

\newpage
\begin{theorem}\label{t31o}
	If $m_2\ge 3$ and $m_2$ is odd, then $\mathcal{C}_{S\times D_3}$ is a ${\Big{[}\frac{(q^{m_1}-1)(q^{m_2-1}-1)}{q-1},m_1+m_2\Big{]}}$ 
	linear code with weight distribution in Table $6$.
	
	\begin{center} Table $6$~~~ The weight distribution of  $\mathcal{C}_{S\times {D}_3}$~$(2\nmid m_2)$
		
		\begin{tabular}{|p{7cm}<{\centering}| p{5cm}<{\centering}|}
	\hline   weight $w$	                     &   frequency $A_w$                    \\ 
	\hline       $0$	                     &  $1$                                    \\
	\hline   $q^{m_1+m_2-2}-q^{m_1-1}$   &  $q^{m_1}-1$        \\ 
	\hline   $q^{m_1+m_2-2}-q^{m_2-2}$   &  $q^{m_2-1}-1$        \\ 
	\hline   $q^{m_1+m_2-2}-q^{m_2-2}-(q^{m_1}-1)q^{\frac{m_2-3}{2}}$	                     &      $\frac{(q-1)}{2}\big(q^{m_2-1}+q^{\frac{m_2-1}{2}}\big)$  \\
	\hline   $q^{m_1+m_2-2}-q^{m_2-2}+(q^{m_1}-1)q^{\frac{m_2-3}{2}}$	                     &      $\frac{(q-1)}{2}\big(q^{m_2-1}-q^{\frac{m_2-1}{2}}\big)$  \\
	\hline   $q^{m_1+m_2-2}-q^{m_2-2}-q^{m_1-1}$   &  $(q^{m_1}-1)(q^{m_2-1}-1)$        \\          
	\hline   $q^{m_1+m_2-2}-q^{m_2-2}-q^{m_1-1}+q^{\frac{m_2-3}{2}}$   &  $ \frac{(q^{m_1}-1)(q-1)}{2}\big(q^{m_2-1}+q^{\frac{m_2-1}{2}}\big)$        \\   
	\hline   $q^{m_1+m_2-2}-q^{m_2-2}-q^{m_1-1}-q^{\frac{m_2-3}{2}}$   &  $ \frac{(q^{m_1}-1)(q-1)}{2}\big(q^{m_2-1}-q^{\frac{m_2-1}{2}}\big)$       \\   
	\hline
\end{tabular} 
	\end{center}
\end{theorem}

\begin{theorem}\label{t32e}
	If $m_2$ is even, then $\mathcal{C}_{S\times \tilde{D}_3}$ is a ${\Big{[}(q^{m_1}-1)(\frac{q^{m_2-1}}{q-1}+\frac{G_{m_2}}{q}),m_1+m_2\Big{]}}$ 
	linear code with weight distribution in Table $7$.
	\begin{center} Table $7$~~~ The weight distribution of  $\mathcal{C}_{S \times {\tilde{D}}_3}$~$(2\mid m_2)$
		
		\begin{tabular}{|p{7.5cm}<{\centering}| p{6cm}<{\centering}|}
			\hline   weight $w$	                     &   frequency $A_w$                    \\ 
			\hline       $0$	                     &  $1$                                    \\
			\hline   $q^{m_1+m_2-2}+(q-1)q^{m_1-2}G_{m_2}$   &  $q^{m_1}-1$        \\ 
			\hline   $q^{m_1+m_2-2}-q^{m_2-2}$   &  $q^{m_2-1}-1+(q-1)\frac{G_{m_2}}{q}$        \\ 
			\hline   $q^{m_1+m_2-2}-q^{m_2-2}+(q^{m_1}-1)\frac{G_{m_2}}{q}$	                     &      $(q-1)\big(q^{m_2-1}+\frac{G_{m_2}}{q}\big)$  \\
			\hline   $q^{m_1+m_2-2}-q^{m_2-2}+(q-1)q^{m_1-2}G_{m_2}$   &  $(q^{m_1}-1)\big(q^{m_2-1}-1+(q-1)\frac{G_{m_2}}{q}\big)$        \\   
			\hline   $q^{m_1+m_2-2}-q^{m_2-2}+\big((q-1)q^{m_1-1}-1\big)\frac{G_{m_2}}{q}$   &  $(q^{m_1}-1)(q-1)\big(q^{m_2-1}+\frac{G_{m_2}}{q}\big)$        \\          
			\hline
		\end{tabular} 
	\end{center}
\end{theorem}

\begin{theorem}\label{t32o}
	If $m_2\ge 3$ and $m_2$ is odd, then $\mathcal{C}_{S\times \tilde{D}_3}$ is a ${\Big{[}\frac{(q^{m_1}-1)q^{m_2-1}}{q-1},m_1+m_2\Big{]}}$ 
	linear code with weight distribution in Table $8$.
	\begin{center} Table $8$~~~ The weight distribution of  $\mathcal{C}_{S \times {\tilde{D}}_3}$~$(2\nmid m_2)$
		
		\begin{tabular}{|p{7cm}<{\centering}| p{5cm}<{\centering}|}
		\hline   weight $w$	                     &   frequency $A_w$                    \\ 
		\hline       $0$	                     &  $1$                                    \\
		\hline   $q^{m_1+m_2-2}$   &  $q^{m_1}-1$        \\ 
		\hline   $q^{m_1+m_2-2}-q^{m_2-2}$   &  $q^{m_1}(q^{m_2-1}-1)$        \\ 
		\hline   $q^{m_1+m_2-2}-q^{m_2-2}-(q^{m_1}-1)q^{\frac{m_2-3}{2}}$	                     &      $\frac{(q-1)}{2}\big(q^{m_2-1}+q^{\frac{m_2-1}{2}}\big)$  \\
		\hline   $q^{m_1+m_2-2}-q^{m_2-2}+(q^{m_1}-1)q^{\frac{m_2-3}{2}}$	                     &      $\frac{(q-1)}{2}\big(q^{m_2-1}-q^{\frac{m_2-1}{2}}\big)$  \\
		\hline   $q^{m_1+m_2-2}-q^{m_2-2}+q^{\frac{m_2-3}{2}}$   &  $ \frac{(q^{m_1}-1)(q-1)}{2}\big(q^{m_2-1}+q^{\frac{m_2-1}{2}}\big)$        \\   
		\hline   $q^{m_1+m_2-2}-q^{m_2-2}-q^{\frac{m_2-3}{2}}$   &  $ \frac{(q^{m_1}-1)(q-1)}{2}\big(q^{m_2-1}-q^{\frac{m_2-1}{2}}\big)$       \\   
		\hline
	\end{tabular} 
	\end{center}
\end{theorem}

\section{Proofs of main results}
\subsection{Some auxiliary lemmas}

\begin{lemma}\label{key0}
	For any $(a,b)\in\mathbb{F}_{q^{m_1}}\times\mathbb{F}_{q^{m_2}}$, let $wt(\mathbf{c}(a,b))$ be the Hamming weight of $\mathbf{c}(a,b)=\big(\mathrm{Tr}^{q^{m_1}}_q(ax)+\mathrm{Tr}^{q^{m_2}}_q(by)\big)_{(x,y)\in \mathrm{S}\times D}$. If  $S$ and $D$ are given in $(\ref{S})$ and $(\ref{D})$, respectively, then\begin{align*}
	wt(\mathbf{c}(a,b))=&\begin{cases}
	0,&\text{ if~}a=b=0;\\
	q^{m_1-1}|D|,&\text{ if~}a\neq 0\text{~and~}b=0;\\
	\frac{(q^{m_1}-1)(|D|-T(D,b))}{q},&\text{ if~}a=0\text{~and~}b\neq 0;\\
	\frac{(q^{m_1}-1)|D|+T(D,b)}{q},&\text{ if~}a\neq 0\text{~and~}b\neq 0,
	\end{cases}
	\end{align*} 
	where
	\begin{align*}
		T(D,b)=\sum\limits_{y\in D}\zeta_p^{\mathrm{Tr}^{q^{m_2}}_p(by)}.
	\end{align*}
\end{lemma}

{\bf Proof}. For any $(a,b)\in\mathbb{F}_{q^{m_1}}\times\mathbb{F}_{q^{m_2}}$, 
let\begin{align*}
N(a,b)=\{(x,y)\in\mathrm{S}\times D~\Big|~\mathrm{Tr}^{q^{m_1}}_q(ax)+\mathrm{Tr}^{q^{m_2}}_q(by)=0\}.
\end{align*}
Note that $\bigcup\limits_{z\in\mathbb{F}_q^{*}}zS=\mathbb{F}_{q^{m_1}}^{*}$ and $zD=D$ $(\forall z\in\mathbb{F}_q^{*})$,
by the orthogonal property for the additive character,  we have
\begin{align*}
	\big|N(a,b)\big|=& \sum\limits_{x\in S}\sum\limits_{y\in D}
	\Big(q^{-1}\sum\limits_{z\in\mathbb{F}_q}\zeta_p^{\mathrm{Tr}^{q}_{p}\big(z\big(\mathrm{Tr}^{q^{m_1}}_q(ax)+\mathrm{Tr}^{q^{m_2}}_q(by)\big)\big)}\Big)\\
	=&q^{-1}\bigg(|S\times D|+\sum\limits_{z\in\mathbb{F}_q^{*}}\sum\limits_{x\in zS}\zeta_p^{\mathrm{Tr}^{q^{m_1}}_p(ax)}\sum\limits_{y\in zD}\zeta_p^{\mathrm{Tr}^{q^{m_2}}_p(by)}\bigg)\\
	=&\frac{|S\times D|}{q}+\frac{1}{q}\sum\limits_{x\in \mathbb{F}_{q^{m_1}}^{*}}\zeta_p^{\mathrm{Tr}^{q^{m_1}}_p(ax)}\sum\limits_{y\in D}\zeta_p^{\mathrm{Tr}^{q^{m_2}}_p(by)}\\
    =&\frac{|S\times D|}{q}+\frac{1}{q}\sum\limits_{x\in \mathbb{F}_{q^{m_1}}^{*}}\zeta_p^{\mathrm{Tr}^{q^{m_1}}_p(ax)}\sum\limits_{y\in D}\zeta_p^{\mathrm{Tr}^{q^{m_2}}_p(by)}\\
    =&\begin{cases}
    |S\times D|,&\text{ if~}a=b=0;\\
    \frac{(|S|-1)\times |D|}{q},&\text{ if~}a\neq 0\text{~and~}b=0;\\
    \frac{|S\times D|}{q}+\frac{(q^{m_1}-1)}{q}\sum\limits_{y\in D}\zeta_p^{\mathrm{Tr}^{q^{m_2}}_p(by)},&\text{ if~}a=0\text{~and~}b\neq 0;\\
    \frac{|S\times D|}{q}-\frac{1}{q}\sum\limits_{y\in D}\zeta_p^{\mathrm{Tr}^{q^{m_2}}_p(by)},&\text{ if~}a\neq 0\text{~and~}b\neq 0.
    \end{cases}
\end{align*}
Note that $|S|=\frac{q^{m_1}-1}{q-1}$, we have
\begin{align*}
wt(\mathbf{c}(a,b))=|S\times D|-\big|N(a,b)\big|=&\begin{cases}
0,&\text{ if~}a=b=0;\\
q^{m_1-1}|D|,&\text{ if~}a\neq 0\text{~and~}b=0;\\
\frac{(q^{m_1}-1)(|D|-T(D,b))}{q},&\text{ if~}a=0\text{~and~}b\neq 0;\\
\frac{(q^{m_1}-1)|D|+T(D,b)}{q},&\text{ if~}a\neq 0\text{~and~}b\neq 0.
\end{cases}
\end{align*}$\hfill\Box$

\begin{lemma}\label{D0}
For $D\subseteq\mathbb{F}_{{q}^{m_2}}^{*}$ with $zD=D~(\forall z\in\mathbb{F}_{q}^{*})$, let $\tilde{D}=D\cup\{0\}$, then
\begin{align*}
	z\tilde{D}=\tilde{D}~(\forall z\in\mathbb{F}_{q}^{*}),\quad|\tilde{D}|=|D|+1,~~\text{and}~~T(\tilde{D},b)=T(D,b)+1.
\end{align*}
\end{lemma}

{\bf Proof}. For any $z\in\mathbb{F}_{q}^{*}$, we have $z\tilde{D}=zD\cup z\{0\}=D\cup\{0\}=\tilde{D}$. Furthermore, it is easy to see that
\begin{align*}
|\tilde{D}|=|D|+1~~\text{and}~~T(\tilde{D},b)=\sum\limits_{y\in D\cup\{0\}}\zeta_p^{\mathrm{Tr}^{q^{m_2}}_p(by)}=T(D,b)+1.
\end{align*}
$\hfill\Box$

\begin{lemma}\label{key1}
	For $b\in\mathbb{F}_{q^{m_2}}^{*}$, let $D_1=\mathbb{F}_{q^{m_2}}^{*}$ and $\tilde{D}_1=\mathbb{F}_{q^{m_2}}$, then the following two assertions hold.
	
	$(1)$  $zD_1=D_1$~$(\forall z\in\mathbb{F}_q^{*})$, $|D_1|=q^{m_2}-1$, and $T(D_1,b)=-1$.
	
	$(2)$  $z\tilde{D}_1=\tilde{D}_1$~$(\forall z\in\mathbb{F}_q^{*})$, $|\tilde{D}_1|=q^{m_2}$, and $T(D_1,b)=0$.	
\end{lemma}

{\bf  Proof}. 
 It is obvious that $zD_1=D_1$~$(\forall z\in\mathbb{F}_q^{*})$ and  $|D_1|=q^{m_2}-1$. Furthermore, by the  orthogonal property for the additive character, we have  $T(D_1,b)=-1$.
Thus $(1)$ holds. Now by $(1)$ and Lemma \ref{D0}, $(2)$ is true. 

$\hfill\Box$

\begin{lemma}\label{key2}
	For $m_2\ge 2$ and $b\in\mathbb{F}_{q^{m_2}}^{*}$, let $D_2=\big\{y\in\mathbb{F}_{q^{m_2}}^{*}|\mathrm{Tr}^{q^{m_2}}_q(y)\neq 0\big\}$ and $\tilde{D}_2=D_2\cup\{0\}$, then the following two assertions hold.

	$(1)$ $zD_2=D_2~(\forall z\in\mathbb{F}_q^{*})$, $|D_2|=(q-1)q^{m_2-1}$, and
	\begin{align*}
	T(D_2,b)=\begin{cases}
	0,&\text{if~}b\in\mathbb{F}_{q^{m_2}}\backslash\mathbb{F}_q;\\
	-q^{m_2-1},&\text{if~}b\in\mathbb{F}_q^{*}.
	\end{cases}
	\end{align*}

	$(2)$ $z\tilde{D}_2=\tilde{D}_2~(\forall z\in\mathbb{F}_q^{*})$, $|\tilde{D}_2|=(q-1)q^{m_2-1}+1$, and
	\begin{align*}
	T(\tilde{D}_2,b)=\begin{cases}
	1,&\text{if~}b\in\mathbb{F}_{q^{m_2}}\backslash\mathbb{F}_q;\\
	1-q^{m_2-1},&\text{if~}b\in\mathbb{F}_q^{*}.
	\end{cases}
	\end{align*}
	
\end{lemma}

{\bf  Proof}.  For any $z\in\mathbb{F}_q^{*}$ and $y\in\mathbb{F}_{q^{m_2}}^{*}$, $\mathrm{Tr}^{q^{m_2}}_q(y)\neq 0$ if and only if $\mathrm{Tr}^{q^{m_2}}_q(zy)\neq 0$, namely,   $zD_2=D_2~(\forall z\in\mathbb{F}_q^{*})$.
Furthermore, by  the  orthogonal property for the additive character, we have
\begin{align*}
	|D_2|=(q-1)q^{m_2-1},
\end{align*}
and
\begin{align*}
T(D_2,b)&=\sum\limits_{y\in \mathbb{F}_{q^{m_2}}}\zeta_p^{\mathrm{Tr}^{q^{m_2}}_p(by)}-\sum\limits_{y\in \mathbb{F}_{q^{m_2}}\backslash D_2}\zeta_p^{\mathrm{Tr}^{q^{m_2}}_p(by)}\\
&=-\sum\limits_{y\in \mathbb{F}_{q^{m_2}}}\Bigg(\frac{1}{q}\sum\limits_{\lambda \in \mathbb{F}_{q}}\zeta_p^{\mathrm{Tr}^{q}_{p}\big(\lambda\mathrm{Tr}^{q^{m_2}}_q(y)\big)}\Bigg)\zeta_p^{\mathrm{Tr}^{q^{m_2}}_p(by)}\\
&=\frac{1}{q}\sum\limits_{\lambda \in \mathbb{F}_{q}}\sum\limits_{y\in \mathbb{F}_{q^{m_2}}}\zeta_p^{\mathrm{Tr}^{q^{m_2}}_p((\lambda+b)y)}\\
&=\begin{cases}
0&\text{if~}b\in\mathbb{F}_{q^{m_2}}^{*}\backslash\mathbb{F}_q^{*};\\
-q^{m_2-1},&\text{if~}b\in\mathbb{F}_q^{*}.
\end{cases}
\end{align*}
So far, $(1)$ holds. Now by $(1)$ and Lemma \ref{D0}, we get $(2)$. 
$\hfill\Box$

\begin{lemma}\label{key3}
	For odd $q$ and $b\in\mathbb{F}_{q^{m_2}}^{*}$, let $D_3=\big\{y\in\mathbb{F}_{q^{m_2}}^{*}~|~\mathrm{Tr}^{q^{m_2}}_q(y^2)= 0\big\}$, then $zD_3=D_3$~$(\forall z\in\mathbb{F}_q^{*})$. Furthermore, if $m_2$ is even, then
	\begin{align*}
	|D_3|=q^{m_2-1}-1+\frac{(q-1)G_{m_2}}{q},
	\end{align*}and
	\begin{align*}
	T(D_3,b)	=\begin{cases}
	-1+\frac{(q-1)G_m}{q},&\text{ if~}\mathrm{Tr}^{q^{m_2}}_{q}(b^2)=0;\\
	-1-\frac{G_m}{q},&\text{ if~}\mathrm{Tr}^{q^{m_2}}_{q}(b^2)\neq 0.
	\end{cases}
	\end{align*}
	If $m_2$ is odd, then
	\begin{align*}
	|D_3|=q^{m_2-1}-1,
	\end{align*}and
	\begin{align*}
	T(D_3,b)=\begin{cases}
	-1,&\text{ if~}\mathrm{Tr}^{q^{m_2}}_{q}(b^2)=0;\\
	-1+\frac{\eta_1\left(-\mathrm{Tr}^{q^{m_2}}_{q}(b^2)\right)G_1G_m}{q},&\text{ if~}\mathrm{Tr}^{q^{m_2}}_{q}(b^2)\neq0.
	\end{cases}
	\end{align*}	
\end{lemma}

{\bf  Proof}.  For any $z\in\mathbb{F}_q^{*}$ and $y\in\mathbb{F}_{q^{m_2}}^{*}$,$\!$ $\mathrm{Tr}^{q^{m_2}}_q(y^2)= 0$ if and only if $\mathrm{Tr}^{q^{m_2}}_q((zy)^2)= 0$, namely, $zD_3=D_3~(\forall z\in\mathbb{F}_q^{*})$.
Furthermore, by  the  orthogonal property for the additive character, Lemmas \ref{l21} and \ref{l28}, we have
\begin{align*}
|D_3|=&\sum\limits_{y\in\mathbb{F}_{q^{m_2}}^{*}}\bigg(\frac{1}{q}\sum\limits_{z \in \mathbb{F}_{q}}\zeta_p^{\mathrm{Tr}^{q}_p\big(z\mathrm{Tr}^{q^{m_2}}_{q}(y^2)\big)}\bigg)\\
=&\frac{q^{m_2}-1}{q}+\frac{1}{q}\sum\limits_{z\in\mathbb{F}_q^{*}}\sum\limits_{y\in\mathbb{F}_{q^{m_2}}^{*}}\zeta_p^{\mathrm{Tr}^{q^{m_2}}_{p}(zy^2)}\\
=&q^{m_2-1}-1+\frac{1}{q}\sum\limits_{z\in\mathbb{F}_q^{*}}\sum\limits_{y\in\mathbb{F}_{q^{m_2}}}\zeta_p^{\mathrm{Tr}^{q^{m_2}}_{p}(zy^2)}\\
=&q^{m_2-1}-1+\frac{G_{m_2}}{q}\sum\limits_{z\in\mathbb{F}_q^{*}}\eta_{m_2}(z)\\
=&\begin{cases}
q^{m_2-1}-1+\frac{(q-1)G_{m_2}}{q},&\text{ if~$m_2$ is even};\\
q^{m_2-1}-1,&\text{ if~$m_2$ is odd},
\end{cases}
\end{align*}
and
\begin{align*}
T(D_3,b)
=&\sum\limits_{y\in \mathbb{F}_{q^{m_2}}^{*}}\bigg(\frac{1}{q}\sum\limits_{z \in \mathbb{F}_{q}}\zeta_p^{\mathrm{Tr}^{q}_p\big(z\mathrm{Tr}^{q^{m_2}}_{q}(y^2)\big)}\bigg)\zeta_p^{\mathrm{Tr}^{q^{m_2}}_p(by)}\\
=&\frac{1}{q}\sum\limits_{y\in \mathbb{F}_{q^{m_2}}^{*}}\bigg(1+\sum\limits_{z \in \mathbb{F}_{q}^{*}}\zeta_p^{\mathrm{Tr}^{q}_p\big(z\mathrm{Tr}^{q^{m_2}}_{q}(y^2)\big)}\bigg)\zeta_p^{\mathrm{Tr}^{q^{m_2}}_p(by)}\\
=&\frac{1}{q}\sum\limits_{y\in \mathbb{F}_{q^{m_2}}^{*}}\zeta_p^{\mathrm{Tr}^{q^{m_2}}_p(by)}+\frac{1}{q}\sum\limits_{z\in\mathbb{F}_q^{*}}\sum\limits_{y\in\mathbb{F}_{q^{m_2}}^{*}}\zeta_p^{\mathrm{Tr}^{q^{m_2}}_{p}(zy^2+by)}\\
=&-\frac{1}{q}-\frac{q-1}{q}+\frac{1}{q}\sum\limits_{z\in\mathbb{F}_q^{*}}\sum\limits_{y\in\mathbb{F}_{q^{m_2}}}\zeta_p^{\mathrm{Tr}^{q^{m_2}}_{p}(zy^2+by)}\\
=&-1+\frac{G_m}{q}\sum\limits_{z\in\mathbb{F}_q^{*}}\eta_{m_2}(z)\zeta_p^{\mathrm{Tr}^{q^{m_2}}_{p}\big(\frac{b^2}{-4z}\big)}.
\end{align*}
Basing on Lemma \ref{l21}, we have the following two cases.

If $m_2$ is even, then
\begin{align*}
T(D_3,b)=&-1+\frac{G_m}{q}\sum\limits_{z\in\mathbb{F}_q^{*}}\zeta_p^{\mathrm{Tr}^{q^{m_2}}_{p}\big(\frac{b^2}{-4z}\big)}\\
=&-1+\frac{G_m}{q}\sum\limits_{z\in\mathbb{F}_q^{*}}\zeta_p^{\mathrm{Tr}^{q}_p\big(\mathrm{Tr}^{q^{m_2}}_{q}(b^2)z\big)}\\
=&\begin{cases}
-1+\frac{(q-1)G_m}{q},&\text{ if~}\mathrm{Tr}^{q^{m_2}}_{q}(b^2)=0;\\
-1-\frac{G_m}{q},&\text{ if~}\mathrm{Tr}^{q^{m_2}}_{q}(b^2)\neq 0.
\end{cases}
\end{align*}

If $m_2$ is odd, then
\begin{align*}
T(D_3,b)=&-1+\frac{G_m}{q}\sum\limits_{z\in\mathbb{F}_q^{*}}\eta_1(z)\zeta_p^{\mathrm{Tr}^{q^{m_2}}_{p}\big(\frac{b^2}{-4z}\big)}\\
=&-1+\frac{G_m}{q}\sum\limits_{z\in\mathbb{F}_q^{*}}\eta_1(z)\zeta_p^{\mathrm{Tr}^{q}_p\big(-\mathrm{Tr}^{q^{m_2}}_{q}(b^2)z\big)}\\
=&\begin{cases}
-1,&\text{ if~}\mathrm{Tr}^{q^{m_2}}_{q}(b^2)=0;\\
-1+\frac{\eta_1\big(-\mathrm{Tr}^{q^{m_2}}_{q}(b^2)\big)G_1G_m}{q},&\text{ if~}\mathrm{Tr}^{q^{m_2}}_{q}(b^2)\neq0.
\end{cases}
\end{align*}

$\hfill\Box$

By Lemmas \ref{D0} and \ref{key3}, we can get the following lemma directly. 
\begin{lemma}\label{key30}
	For odd $q$ and $b\in\mathbb{F}_{q^{m_2}}^{*}$, let $\tilde{D}_3=D_3\cup \{0\}$, then $z\tilde{D}_3=\tilde{D}_3$~$(\forall z\in\mathbb{F}_q^{*})$. Furthermore, if $m_2$ is even, then
	\begin{align*}
	|\tilde{D}_3|=q^{m_2-1}+\frac{(q-1)G_{m_2}}{q},
	\end{align*}and
	\begin{align*}
	T(\tilde{D}_3,b)	=\begin{cases}
	\frac{(q-1)G_m}{q},&\text{ if~}\mathrm{Tr}^{q^{m_2}}_{q}(b^2)=0;\\
	-\frac{G_m}{q},&\text{ if~}\mathrm{Tr}^{q^{m_2}}_{q}(b^2)\neq 0.
	\end{cases}
	\end{align*}
	If $m_2$ is odd, then
	\begin{align*}
	|\tilde{D}_3|=q^{m_2-1},
	\end{align*}and
	\begin{align*}
	T(\tilde{D}_3,b)=\begin{cases}
	0,&\text{ if~}\mathrm{Tr}^{q^{m_2}}_{q}(b^2)=0;\\
	\frac{\eta_1\big(-\mathrm{Tr}^{q^{m_2}}_{q}(b^2)\big)G_1G_m}{q},&\text{ if~}\mathrm{Tr}^{q^{m_2}}_{q}(b^2)\neq0.
	\end{cases}
	\end{align*}	
\end{lemma}

\begin{lemma}
	For $\rho\in\mathbb{F}_{q}$, let $N_\rho=\big\{b\in\mathbb{F}_{q^{m_2}}^{*}~|~\mathrm{Tr}^{q^{m_2}}_q(b^2)= \rho\big\}$, then the following two assertions hold.
	
	$(1)$ If $m_2$ is even, then
	\begin{align*}
	|N_\rho|=\begin{cases}
	q^{m_2-1}-1+\frac{(q-1)G_{m_2}}{q},&\text{if~}\rho=0;\\
	q^{m_2-1}-\frac{G_{m_2}}{q},&\text{if~}\rho\neq 0.
	\end{cases}
	\end{align*}
	
	$(2)$ If $m_2$ is even, then
	\begin{align*}
	|N_\rho|=\begin{cases}
	q^{m_2-1}-1,&\text{if~}\rho=0;\\
	q^{m_2-1}+\frac{\eta_{1}(-\rho)G_{m_2}G_1}{q},&\text{if~}\rho\neq 0.
	\end{cases}
	\end{align*}
\end{lemma}

{\bf Proof}. 
For $\rho=0$, note that $N_{\rho}=D_3$, thus we can get  $|N_{\rho}|$ from Lemma \ref{key3}.

For $\rho\neq 0$, by  the  orthogonal property for the additive character and Lemmas \ref{l21} and \ref{l28}, we have
\begin{align*}
|N_{\rho}|=&\sum\limits_{b\in\mathbb{F}_{q^{m_2}}^{*}}\bigg(\frac{1}{q}\sum\limits_{z \in \mathbb{F}_{q}}\zeta_p^{\mathrm{Tr}^{q}_p\big(z(\mathrm{Tr}^{q^{m_2}}_{q}(b^2)-\rho)\big)}\bigg)\\
=&\frac{q^{m_2}-1}{q}+\frac{1}{q}\sum\limits_{z\in\mathbb{F}_q^{*}}\zeta_p^{\mathrm{Tr}^{q}_p(-\rho z)}\sum\limits_{b\in\mathbb{F}_{q^{m_2}}^{*}}\zeta_p^{\mathrm{Tr}^{q^{m_2}}_{p}(zb^2)}\\
=&q^{m_2-1}+\frac{1}{q}\sum\limits_{z\in\mathbb{F}_q^{*}}\zeta_p^{\mathrm{Tr}^{q}_p(-\rho z)}\sum\limits_{y\in\mathbb{F}_{q^{m_2}}}\zeta_p^{\mathrm{Tr}^{q^{m_2}}_{p}(zb^2)}\\
=&q^{m_2-1}+\frac{G_{m_2}}{q}\sum\limits_{z\in\mathbb{F}_q^{*}}\zeta_p^{\mathrm{Tr}^{q}_p(-\rho z)}\eta_{m_2}(z)\\
=&\begin{cases}
q^{m_2-1}-\frac{G_{m_2}}{q},&\text{ if~$m_2$ is even};\\
q^{m_2-1}+\frac{\eta_{1}(-\rho)G_{m_1}G_1}{q},&\text{ if~$m_2$ is odd}.
\end{cases}
\end{align*}
$\hfill\Box$

\subsection{Proofs for Theorems \ref{t1}-\ref{t32o}}

In the following, for any  $\mathbf{c}(a,b)=\big(\mathrm{Tr}^{q^{m_1}}_q(ax)+\mathrm{Tr}^{q^{m_2}}_q(by)\big)_{(x,y)\in \mathrm{S}\times D}\in\mathcal{C}_{S\times D}$, let $A_{wt(\mathbf{c}(a,b))}$ be the number of codewords in $\mathcal{C}_{S\times D}$ with Hamming weight $wt(\mathbf{c}(a,b))$.

{\bf Proofs for Theorem \ref{t1}}.  For $D=D_1=\mathbb{F}_{q^{m_2}}^{*}$, by Lemmas \ref{key0} and \ref{key1}, the length of $\mathcal{C}_{S\times D_1}$ is $\frac{(q^{m_1}-1)(q^{m_2}-1)}{q-1}$.
Furthermore, 
\begin{align*}
wt(\mathbf{c}(a,b))=&\begin{cases}
0,&\text{ if~}a=b=0;\\
q^{m_1+m_2-1}-q^{m_1-1},&\text{ if~}a\neq 0\text{~and~}b=0;\\
q^{m_1+m_2-1}-q^{m_2-1},&\text{ if~}a=0\text{~and~}b\neq 0;\\
q^{m_1+m_2-1}-q^{m_1-1}-q^{m_2-1},&\text{ if~}a\neq 0\text{~and~}b\neq 0.
\end{cases}
\end{align*}
Thus,\begin{align*}
	&A_{q^{m_1+m_2-1}-q^{m_1-1}}=q^{m_1}-1;\\
	&A_{q^{m_1+m_2-1}-q^{m_2-1}}=q^{m_2}-1;\\
	&A_{q^{m_1+m_2-1}-q^{m_1-1}-q^{m_2-1}}=(q^{m_1}-1)(q^{m_2}-1).
\end{align*}
By the above disscussions,   $\mathcal{C}_{S\times D_1}$ is a $\left[\frac{(q^{m_1}-1)(q^{m_2}-1)}{q-1},m_1+m_2,q^{m_1+m_2-1}-q^{m_1-1}-q^{m_2-1}\right]$ linear code with weight distribution in Table $1$. 

Now we show that $\mathcal{C}_{S\times D_1}$ is a near Griesmer or Griesmer code by the following two cases.

For $m_1=m_2$, we have\begin{align*}
&\sum\limits_{i=0}^{m_1+m_2-1}\left\lceil\frac{q^{m_1+m_2-1}-q^{m_1-1}-q^{m_2-1}}{q^{i}}\right\rceil\\
=&\begin{cases}
\sum\limits_{i=0}^{m_1}\big(q^{2m_1-1-i}-2q^{m_1-1-i}\big)+\sum\limits_{i=m_1+1}^{2m_1-1}q^{2m_1-1-i},&\quad\text{ if~}q=2;\\
\sum\limits_{i=0}^{m_1-1}\big(q^{2m_1-1-i}-2q^{m_1-1-i}\big)+\sum\limits_{i=m_1}^{2m_1-1}q^{2m_1-1-i},&\quad\text{ if~}q\neq 2;
\end{cases}\\
=&\begin{cases}
(2^{m_1}-1)^2-1,&\quad\text{ if~}q=2;\\
\frac{(q^{m_1}-1)^2}{q-1},&\quad\text{ if~}q\neq 2.
\end{cases}
\end{align*}

For $m_1\neq m_2$, let $\bar{m}=\min\{m_1,m_2\}$ and $\tilde{m}=\max\{m_1,m_2\}$, we have
\begin{align*}
	&\sum\limits_{i=0}^{m_1+m_2-1}\left\lceil\frac{q^{m_1+m_2-1}-q^{m_1-1}-q^{m_2-1}}{q^{i}}\right\rceil\\
	=&\sum\limits_{i=0}^{\bar{m}+\tilde{m}-1}\left\lceil\frac{q^{\bar{m}+\tilde{m}-1}-q^{\bar{m}-1}-q^{\tilde{m}-1}}{q^{i}}\right\rceil\\
	=&	\sum\limits_{i=0}^{\bar{m}-1}\big(q^{\bar{m}+\tilde{m}-1-i}-q^{\bar{m}-1-i}-q^{\tilde{m}-1-i}\big)+	\sum\limits_{i=\bar{m}}^{\tilde{m}-1}\big(q^{\bar{m}+\tilde{m}-1-i}-q^{\tilde{m}-1-i}\big)+\sum\limits_{i=\tilde{m}}^{\bar{m}+\tilde{m}-1}q^{\bar{m}+\tilde{m}-1-i}\\
	=&\frac{q^{\bar{m}+\tilde{m}}-1}{q-1}-\frac{q^{\bar{m}}-1}{q-1}-\frac{q^{\tilde{m}}-1}{q-1}\\
	=&\frac{(q^{m_1}-1)(q^{m_2}-1)}{q-1}.
\end{align*}

Now by (\ref{GB}) and $\sum\limits_{i=0}^{2m_1-1}\left\lceil\frac{2^{2m_1-1}-2^{m_1}+1}{2^{i}}\right\rceil>(2^{m_1}-1)^2,$  $\mathcal{C}_{S\times D_1}$ is a distance-optimal near Griesmer code if $m_1=m_2$ and $q=2$. Otherwise,  $\mathcal{C}_{S\times D_1}$ is a Griesmer code.
$\hfill\Box$\\

{\bf Proofs for Theorem \ref{t11}}.  For $D=\tilde{D}_1=\mathbb{F}_{q^{m_2}}$, by Lemmas \ref{key0} and \ref{key1}, the length of $\mathcal{C}_{\tilde{D}_1}$ is $\frac{(q^{m_1}-1)q^{m_2}}{q-1}$. Furthermore, 
\begin{align*}
wt(\mathbf{c}(a,b))=&\begin{cases}
0,&\text{ if~}a=b=0;\\
q^{m_1+m_2-1},&\text{ if~}a\neq 0\text{~and~}b=0;\\
q^{m_1+m_2-1}-q^{m_2-1},&\text{ if~}b\neq 0.
\end{cases}
\end{align*}
Thus,
\begin{align*}
	&A_{q^{m_1+m_2-1}}=q^{m_1}-1,\\
	&A_{q^{m_1+m_2-1}-q^{m_2-1}}=q^{m_1+m_2}-q^{m_1}.
\end{align*}
By the above disscussions,  $\mathcal{C}_{S\times D_1}$ is a $\left[\frac{(q^{m_1}-1)q^{m_2}}{q-1},m_1+m_2,q^{m_1+m_2-1}-q^{m_2-1}\right]$ linear code with weight distribution in Table $2$. 

Note that
\begin{align*}
&\sum\limits_{i=0}^{m_1+m_2-1}\left\lceil\frac{q^{m_1+m_2-1}-q^{m_2-1}}{q^{i}}\right\rceil\\
=&	\sum\limits_{i=0}^{m_2-1}\big(q^{m_1+m_2-1-i}-q^{m_2-1-i}\big)+\sum\limits_{i=m_2}^{m_1+m_2-1}q^{m_1+m_2-1-i}\\
=&\frac{q^{m_1+m_2}-1}{q-1}-\frac{q^{m_2}-1}{q-1}\\
=&\frac{(q^{m_1}-1)q^{m_2}}{q-1},
\end{align*}
by (\ref{GB}), $\mathcal{C}_{S\times D_1}$ is a Griesmer code.
$\hfill\Box$\\

{\bf Proofs for Theorem \ref{t2}}.  For $m_2\ge 2$, $D=D_2=\big\{y\in\mathbb{F}_{q^{m_2}}^{*}|\mathrm{Tr}^{q^{m_2}}_q(y)\neq 0\big\}$, by  Lemmas \ref{key0} and \ref{key2},  the length of $\mathcal{C}_{S\times D_2}$ is $q^{m_1+m_2-1}-q^{m_2-1}$. Furthermore,
\begin{align*}
wt(\mathbf{c}(a,b))=&\begin{cases}
0,&\text{ if~}a=b=0;\\
q^{m_1+m_2-1}-q^{m_1+m_2-2},&\text{ if~}a\neq 0\text{~and~}b=0;\\
q^{m_1+m_2-1}-q^{m_1+m_2-2}-q^{m_2-1}+q^{m_2-2},&\text{ if~}b\in\mathbb{F}_{q^{m_2}}^{*}\backslash\mathbb{F}_{q}^{*};\\
q^{m_1+m_2-1}-q^{m_2-1},&\text{ if~}a=0\text{~and~}b\in\mathbb{F}_{q}^{*};\\
q^{m_1+m_2-1}-q^{m_1+m_2-2}-q^{m_2-1},&\text{ if~}a\neq 0\text{~and~}b\in\mathbb{F}_{q}^{*}.
\end{cases}
\end{align*}
Thus,\begin{align*}
&A_{q^{m_1+m_2-1}-q^{m_1+m_2-2}}=q^{m_1}-1;\\
&A_{q^{m_1+m_2-1}-q^{m_1+m_2-2}-q^{m_2-1}+q^{m_2-2}}=q^{m_1+m_2}-q^{m_1+1};\\
&A_{q^{m_1+m_2-1}-q^{m_2-1}}=q-1;\\
&A_{q^{m_1+m_2-1}-q^{m_1+m_2-2}-q^{m_2-1}}=q^{m_1+1}-q^{m_1}-q+1.
\end{align*}
By the above disscussions,  $\mathcal{C}_{S\times D_2}$ is a $\left[q^{m_1+m_2-1}-q^{m_2-1},m_1+m_2,q^{m_1+m_2-1}-q^{m_1+m_2-2}-q^{m_2-1}\right]$ linear code with weight distribution in Table $3$.

Note that 
\begin{align*}
&\sum\limits_{i=0}^{m_1+m_2-1}\left\lceil\frac{q^{m_1+m_2-1}-q^{m_1+m_2-2}-q^{m_2-1}}{q^{i}}\right\rceil\\
=&	\sum\limits_{i=0}^{m_2-1}\big(q^{m_1+m_2-1-i}-q^{m_1+m_2-2-i}-q^{m_2-1-i}\big)+\sum\limits_{i=m_2}^{m_1+m_2-2}\big(q^{m_1+m_2-1-i}-q^{m_1+m_2-2-i}\big)+1\\
=&\frac{q^{m_1+m_2}-1}{q-1}-\frac{q^{m_1+m_2-1}-1}{q-1}-\frac{q^{m_2}-1}{q-1}\\
=&q^{m_1+m_2-1}-\frac{q^{m_2}-1}{q-1},
\end{align*}
thus we have
\begin{align*}
\sum\limits_{i=0}^{m_1+1}\left\lceil\frac{q^{m_1+1}-q^{m_1}-q}{q^{i}}\right\rceil=q^{m_1+1}-q-1,
\end{align*}
and
\begin{align*}
\sum\limits_{i=0}^{m_1+1}\left\lceil\frac{q^{m_1+1}-q^{m_1}-q+1}{q^{i}}\right\rceil=q^{m_1+1}-q+1.
\end{align*}
Now by (\ref{GB}),  $\mathcal{C}_{S\times D_2}$ is a distance-optimal near Griesmer code for $m_2=2$.$\hfill\Box$\\

{\bf Proof for Theorem \ref{t21}}.  For $m_2\ge 2$, $D=\tilde{D}_2=D_2\cup\{0\}$, by  Lemmas \ref{key0} and \ref{key2},  the length of $\mathcal{C}_{S\times \tilde{D}_2}$ is $q^{m_1+m_2-1}-q^{m_2-1}+\frac{q^{m_1}-1}{q-1}$. Furthermore,
\begin{align*}
wt(\mathbf{c}(a,b))=&\begin{cases}
0,&\!\!\!\!\text{ if~}a=b=0;\\
q^{m_1+m_2-1}-q^{m_1+m_2-2}+q^{m_1-1},&\!\!\!\!\text{ if~}a\neq 0\text{~and~}b=0;\\
q^{m_1+m_2-1}-q^{m_1+m_2-2}-q^{m_2-1}+q^{m_2-2},&\!\!\!\!\text{ if~}a=0\text{~and~}b\in\mathbb{F}_{q^{m_2}}^{*}\backslash\mathbb{F}_{q}^{*};\\
q^{m_1+m_2-1}-q^{m_2-1},&\!\!\!\!\text{ if~}a=0\text{~and~}b\in\mathbb{F}_{q}^{*};\\
q^{m_1+m_2-1}-q^{m_1+m_2-2}+q^{m_1-1}-q^{m_2-1}+q^{m_2-2},&\!\!\!\!\text{ if~}a\neq 0\text{~and~}b\in\mathbb{F}_{q^{m_2}}^{*}\backslash\mathbb{F}_{q}^{*};\\
q^{m_1+m_2-1}-q^{m_1+m_2-2}+q^{m_1-1}-q^{m_2-1},&\!\!\!\!\text{ if~}a\neq 0\text{~and~}b\in\mathbb{F}_{q}^{*}.
\end{cases}
\end{align*}
Thus,\begin{align*}
&A_{q^{m_1+m_2-1}-q^{m_1+m_2-2}+q^{m_1-1}}=q^{m_1}-1;\\
&A_{q^{m_1+m_2-1}-q^{m_1+m_2-2}-q^{m_2-1}+q^{m_2-2}}=q^{m_2}-q;\\
&A_{q^{m_1+m_2-1}-q^{m_2-1}}=q-1;\\
&A_{q^{m_1+m_2-1}-q^{m_1+m_2-2}+q^{m_1-1}-q^{m_2-1}+q^{m_2-2}}=q^{m_1+m_2}-q^{m_1+1}-q^{m_2}+q;\\
&A_{q^{m_1+m_2-1}-q^{m_1+m_2-2}+q^{m_1-1}-q^{m_2-1}}=q^{m_1+1}-q^{m_1}-q+1.
\end{align*}
$\hfill\Box$\\

{\bf Proofs for Theorems \ref{t31e}-\ref{t31o}}. For $m_2\ge 2$ and $D={D}_3$, by  Lemmas \ref{key0} and \ref{key3}, we have the following two cases.

If $m_2$ is even, then the length of $\mathcal{C}_{S\times D_3}$ is $(q^{m_1}-1)(\frac{q^{m_2-1}-1}{q-1}+\frac{G_{m_2}}{q})$. Furthermore,{\small
\begin{align*}
&wt(\mathbf{c}(a,b))\\
=&\begin{cases}
0,&\!\!\!\!\text{ if~}a=b=0;\\
q^{m_1+m_2-2}-q^{m_1-1}+(q-1)q^{m_1-2}G_{m_2},&\!\!\!\!\text{ if~}a\neq 0\text{~and~}b=0;\\
q^{m_1+m_2-2}-q^{m_2-2},&\!\!\!\!\text{ if~}a=0,~b\neq 0\text{~and~}\mathrm{Tr}_{q}^{q^{m_2}}(b^2)=0;\\
q^{m_1+m_2-2}-q^{m_2-2}+(q^{m_1}-1)\frac{G_{m_2}}{q},&\!\!\!\!\text{ if~}a=0\text{~and~}\mathrm{Tr}_{q}^{q^{m_2}}(b^2)\neq 0;\\
q^{m_1+m_2-2}-q^{m_2-2}-q^{m_1-1}+(q-1)q^{m_1-2}G_{m_2},&\!\!\!\!\text{ if~}a\neq 0,~b\neq 0\text{~and~}\mathrm{Tr}_{q}^{q^{m_2}}(b^2)=0;\\
q^{m_1+m_2-2}-q^{m_2-2}-q^{m_1-1}+((q-1)q^{m_1-1}-1)\frac{G_{m_2}}{q},&\!\!\!\!\text{ if~}a\neq 0\text{~and~}\mathrm{Tr}_{q}^{q^{m_2}}(b^2)\neq 0.
\end{cases}
\end{align*} }
Now by Lemma \ref{key30}, we have{\small
\begin{align*}
	&A_{q^{m_1+m_2-2}-q^{m_1-1}+(q-1)q^{m_1-2}G_{m_2}}=q^{m_1}-1;\\
	&A_{q^{m_1+m_2-2}-q^{m_2-2}}=|N_0|=q^{m_2-1}-1+(q-1)\frac{G_{m_2}}{q};\\
	&A_{q^{m_1+m_2-2}-q^{m_2-2}+(q^{m_1}-1)\frac{G_{m_2}}{q}}=\sum\limits_{\rho\in\mathbb{F}_q^{*}}|N_\rho|=(q-1)\big(q^{m_2-1}+\frac{G_{m_2}}{q}\big);\\
	&A_{q^{m_1+m_2-2}-q^{m_2-2}-q^{m_1-1}+(q-1)q^{m_1-2}G_{m_2}}=(q^{m_1}-1)|N_0|=(q^{m_1}-1)\big(q^{m_2-1}-1+(q-1)\frac{G_{m_2}}{q}\big);\\
	&A_{q^{m_1+m_2-2}-q^{m_2-2}-q^{m_1-1}+((q-1)q^{m_1-1}-1)\frac{G_{m_2}}{q}}\!=\!(q^{m_1}-1)\!\!\sum\limits_{\rho\in\mathbb{F}_q^{*}}\!|N_\rho|\!=(q^{m_1}-1)(q-1)\big(q^{m_2-1}\!+\!\frac{G_{m_2}}{q}\big).\\
\end{align*}}

If $m_2$ is odd, then the length of $\mathcal{C}_{S\times D_3}$ is $\frac{(q^{m_1}-1)(q^{m_2-1}-1)}{q-1}$. Furthermore,
\begin{align*}
&wt(\mathbf{c}(a,b))\\
=&\begin{cases}
0,&\text{ if~}a=b=0;\\
q^{m_1+m_2-2}-q^{m_1-1},&\text{ if~}a\neq 0\text{~and~}b=0;\\
q^{m_1+m_2-2}-q^{m_2-2},&\text{ if~}a=0,~b\neq 0\text{~and~}\mathrm{Tr}_{q}^{q^{m_2}}(b^2)=0;\\
q^{m_1+m_2-2}-q^{m_2-2}-(q^{m_1}-1)\frac{G_1G_{m_2}}{q^2},&\text{ if~}a=0\text{~and~}\eta_1\big(-\mathrm{Tr}_{q}^{q^{m_2}}(b^2)\big)= 1;\\
q^{m_1+m_2-2}-q^{m_2-2}+(q^{m_1}-1)\frac{G_1G_{m_2}}{q^2},&\text{ if~}a=0\text{~and~}\eta_1\big(-\mathrm{Tr}_{q}^{q^{m_2}}(b^2)\big)= -1;\\
q^{m_1+m_2-2}-q^{m_2-2}-q^{m_1-1},&\text{ if~}a\neq 0,~b\neq 0\text{~and~}\mathrm{Tr}_{q}^{q^{m_2}}(b^2)=0;\\
q^{m_1+m_2-2}-q^{m_2-2}-q^{m_1-1}+\frac{G_1G_{m_2}}{q^2},&\text{ if~}a\neq 0\text{~and~}\eta_1\big(-\mathrm{Tr}_{q}^{q^{m_2}}(b^2)\big)= 1;\\
q^{m_1+m_2-2}-q^{m_2-2}-q^{m_1-1}-\frac{G_1G_{m_2}}{q^2},&\text{ if~}a\neq 0\text{~and~}\eta_1\big(-\mathrm{Tr}_{q}^{q^{m_2}}(b^2)\big)= -1;\\
\end{cases}
\end{align*} 
Now by Lemma \ref{key30}, we have
\begin{align*}
&A_{q^{m_1+m_2-2}-q^{m_1-1}}=q^{m_1}-1;\\
&A_{q^{m_1+m_2-2}-q^{m_2-2}}=|N_0|=q^{m_2-1}-1;\\
&A_{q^{m_1+m_2-2}-q^{m_2-2}-(q^{m_1}-1)\frac{G_1G_{m_2}}{q^2}}=\sum\limits_{\eta_{1}(-\rho)=1}|N_\rho|=\frac{(q-1)}{2}\big(q^{m_2-1}+\frac{G_1G_{m_2}}{q}\big);\\
&A_{q^{m_1+m_2-2}-q^{m_2-2}+(q^{m_1}-1)\frac{G_1G_{m_2}}{q^2}}=\sum\limits_{\eta_{1}(-\rho)=-1}|N_\rho|=\frac{(q-1)}{2}\big(q^{m_2-1}-\frac{G_1G_{m_2}}{q}\big);\\
&A_{q^{m_1+m_2-2}-q^{m_2-2}-q^{m_1-1}}=(q^{m_1}-1)|N_0|=(q^{m_1}-1)(q^{m_2-1}-1);\\
&A_{q^{m_1+m_2-2}-q^{m_2-2}-q^{m_1-1}+\frac{G_1G_{m_2}}{q^2}}\!\!=(q^{m_1}-1)\!\!\sum\limits_{\eta_{1}(-\rho)=1}\!\!\!|N_\rho|\!=\frac{(q^{m_1}-1)(q-1)}{2}\big(q^{m_2-1}+\!\frac{G_1G_{m_2}}{q}\big);\\
&A_{q^{m_1+m_2-2}-q^{m_2-2}-q^{m_1-1}-\frac{G_1G_{m_2}}{q^2}}\!\!=(q^{m_1}-1)\!\!\sum\limits_{\eta_{1}(-\rho)=-1}\!\!\!|N_\rho|\!=\frac{(q^{m_1}-1)(q-1)}{2}\big(q^{m_2-1}-\!\frac{G_1G_{m_2}}{q}\big).
\end{align*}

By the above disscussions and Lemma \ref{l22}, we can get Theorems \ref{t31e}-\ref{t31o}. 
$\hfill\Box$\\

{\bf Proofs for Theorems \ref{t32e}-\ref{t32o}}. For $m_2\ge 3$ and $D=\tilde{D}_3={D}_3\cup\{0\}$, by  Lemmas \ref{key0} and \ref{key3}, we have the following two cases.

If $m_2$ is even, then the length of $\mathcal{C}_{S\times \tilde{D}_3}$ is $(q^{m_1}-1)(\frac{q^{m_2-1}}{q-1}+\frac{G_{m_2}}{q})$. Furthermore,
\begin{align*}
&wt(\mathbf{c}(a,b))\\
=&\begin{cases}
0,&\text{ if~}a=b=0;\\
q^{m_1+m_2-2}+(q-1)q^{m_1-2}G_{m_2},&\text{ if~}a\neq 0\text{~and~}b=0;\\
q^{m_1+m_2-2}-q^{m_2-2},&\text{ if~}a=0,~b\neq 0\text{~and~}\mathrm{Tr}_{q}^{q^{m_2}}(b^2)=0;\\
q^{m_1+m_2-2}-q^{m_2-2}+(q^{m_1}-1)\frac{G_{m_2}}{q},&\text{ if~}a=0\text{~and~}\mathrm{Tr}_{q}^{q^{m_2}}(b^2)\neq 0;\\
q^{m_1+m_2-2}-q^{m_2-2}+(q-1)q^{m_1-2}G_{m_2},&\text{ if~}a\neq 0,~b\neq 0\text{~and~}\mathrm{Tr}_{q}^{q^{m_2}}(b^2)=0;\\
q^{m_1+m_2-2}-q^{m_2-2}\big((q-1)q^{m_1-1}-1\big)\frac{G_{m_2}}{q},&\text{ if~}a\neq 0\text{~and~}\mathrm{Tr}_{q}^{q^{m_2}}(b^2)\neq 0.
\end{cases}
\end{align*} 
Now by Lemma \ref{key30}, we have
\begin{align*}
&A_{q^{m_1+m_2-2}+(q-1)q^{m_1-2}G_{m_2}}=q^{m_1}-1;\\
&A_{q^{m_1+m_2-2}-q^{m_2-2}}=|N_0|=q^{m_2-1}-1+(q-1)\frac{G_{m_2}}{q};\\
&A_{q^{m_1+m_2-2}-q^{m_2-2}+(q^{m_1}-1)\frac{G_{m_2}}{q}}=\sum\limits_{\rho\in\mathbb{F}_q^{*}}|N_\rho|=(q-1)\big(q^{m_2-1}+\frac{G_{m_2}}{q}\big);\\
&A_{q^{m_1+m_2-2}-q^{m_2-2}+(q-1)q^{m_1-2}G_{m_2}}=(q^{m_1}-1)|N_0|=(q^{m_1}-1)\big(q^{m_2-1}-1+(q-1)\frac{G_{m_2}}{q}\big);\\
&A_{q^{m_1+m_2-2}-q^{m_2-2}+((q-1)q^{m_1-1}-1)\frac{G_{m_2}}{q}}=(q^{m_1}-1)\!\sum\limits_{\rho\in\mathbb{F}_q^{*}}|N_\rho|=(q^{m_1}-1)(q-1)\!\big(q^{m_2-1}\!+\!\frac{G_{m_2}}{q}\big).
\end{align*}

If $m_2$ is odd, then the length of $\mathcal{C}_{S\times \tilde{D}_3}$ is $\frac{(q^{m_1}-1)q^{m_2-1}}{q-1}$. Furthermore,
\begin{align*}
&wt(\mathbf{c}(a,b))\\
=&\begin{cases}
0,&\text{ if~}a=b=0;\\
q^{m_1+m_2-2},&\text{ if~}a\neq 0\text{~and~}b=0;\\
q^{m_1+m_2-2}-q^{m_2-2},&\text{ if~}b\neq 0\text{~and~}\mathrm{Tr}_{q}^{q^{m_2}}(b^2)=0;\\
q^{m_1+m_2-2}-q^{m_2-2}-(q^{m_1}-1)\frac{G_1G_{m_2}}{q^2},&\text{ if~}a=0\text{~and~}\eta_1\big(-\mathrm{Tr}_{q}^{q^{m_2}}(b^2)\big)= 1;\\
q^{m_1+m_2-2}-q^{m_2-2}+(q^{m_1}-1)\frac{G_1G_{m_2}}{q^2},&\text{ if~}a=0\text{~and~}\eta_1\big(-\mathrm{Tr}_{q}^{q^{m_2}}(b^2)\big)= -1;\\
q^{m_1+m_2-2}-q^{m_2-2}+\frac{G_1G_{m_2}}{q^2},&\text{ if~}a\neq 0\text{~and~}\eta_1\big(-\mathrm{Tr}_{q}^{q^{m_2}}(b^2)\big)= 1;\\
q^{m_1+m_2-2}-q^{m_2-2}-\frac{G_1G_{m_2}}{q^2},&\text{ if~}a\neq 0\text{~and~}\eta_1\big(-\mathrm{Tr}_{q}^{q^{m_2}}(b^2)\big)= -1.
\end{cases}
\end{align*} 
Now by Lemma \ref{key30}, we have
\begin{align*}
&A_{q^{m_1+m_2-2}-q^{m_1-1}}=q^{m_1}-1;\\
&A_{q^{m_1+m_2-2}-q^{m_2-2}}=q^{m_1}|N_0|=q^{m_1+m_2-1}-q^{m_1};\\
&A_{q^{m_1+m_2-2}-q^{m_2-2}-(q^{m_1}-1)\frac{G_1G_{m_2}}{q^2}}=\sum\limits_{\eta_{1}(-\rho)=1}|N_\rho|=\frac{(q-1)}{2}\big(q^{m_2-1}+\frac{G_1G_{m_2}}{q}\big);\\
&A_{q^{m_1+m_2-2}-q^{m_2-2}+(q^{m_1}-1)\frac{G_1G_{m_2}}{q^2}}=\sum\limits_{\eta_{1}(-\rho)=-1}|N_\rho|=\frac{(q-1)}{2}\big(q^{m_2-1}-\frac{G_1G_{m_2}}{q}\big);\\
&A_{q^{m_1+m_2-2}-q^{m_2-2}-q^{m_1-1}+\frac{G_1G_{m_2}}{q^2}}\!=\!(q^{m_1}-1)\!\!\sum\limits_{\eta_{1}(-\rho)=1}\!\!|N_\rho|\!=\!\frac{(q^{m_1}-1)(q-1)}{2}\!\big(q^{m_2-1}\!+\!\frac{G_1G_{m_2}}{q}\big);\\
&A_{q^{m_1+m_2-2}-q^{m_2-2}-q^{m_1-1}-\frac{G_1G_{m_2}}{q^2}}\!=\!(q^{m_1}-1)\!\!\sum\limits_{\eta_{1}(-\rho)=-1}\!\!|N_\rho|\!=\!\frac{(q^{m_1}-1)(q-1)}{2}\!\big(q^{m_2-1}\!-\!\frac{G_1G_{m_2}}{q}\big).
\end{align*}

By the above disscussions and Lemma \ref{l22}, we can get Theorems \ref{t32e}-\ref{t32o}. 
$\hfill\Box$\\

\section{Applications}

It is well-known that few-weight linear codes have better applications in secret sharing schemes \cite{JY2006,CC2005}.In particular, projective two-weight codes are very precious as they are closely related to finite projective spaces, strongly regular graphs and combinatorial designs \cite{RC1986,CD2018,P1972}. Here, we present the following two applications.

\subsection{Applications for secret sharing schemes}
The secret sharing schemes is introduced by Blakley \cite{BG} and Shamir \cite{KK} in 1979. Based on linear codes, many secret sharing  schemes are constructed \cite{MJ,JY2006, MJ93, KD2015}. Especially, for those linear codes with any nonzero codewords minimal, their dual codes can be used to construct secret sharing schemes with nice access structures \cite{KD2015}. 

For a linear code $\mathcal{C}$ with length $n$, the support of a codeword $\mathbf{c}=(c_1,\ldots,c_n)\in\mathcal{C}\backslash\{\mathbf{0}\}$ is denoted by 
\begin{align*}
\mathrm{supp}(\mathbf{c})=\{i~|~c_i\neq 0,i=1,\ldots,n\}.
\end{align*} For $\mathbf{c}_1,\mathbf{c}_2\in\mathcal{C}$, when $\mathrm{supp}(\mathbf{c}_2)\subseteq\mathrm{supp}(\mathbf{c}_1)$, we say that $\mathbf{c}_1$ covers $\mathbf{c}_2$ . 
A nonzero codeword $\mathbf{c}\in\mathcal{C}$ is  minimal if it covers only the codeword $\lambda\mathbf{c}$ $(\lambda\in\mathbb{F}_q^{*})$, but no other codewords in $\mathcal{C}$. 

The following lemma is a criteria for the minimal linear code.
\begin{lemma}[Ashikhmin-Barg Lemma \cite{AA}]\label{51}
	Let $w_{min}$ and $w_{max}$ be the minimal and maximal nonzero weights of the linear code $\mathcal{C}$ over $\mathbb{F}_q$, respectively. If
	\begin{align*}
	\frac{w_{min}}{w_{max}}>\frac{q-1}{q},
	\end{align*}
\end{lemma}
then  $\mathcal{C}$ is minimal.\\

Now for the linear code $\mathcal{C}_{D}$ $(D=D_3~\text{or}~\tilde{D}_{3})$, by Theorems $\ref{t31e}$-$\ref{t32o}$,  one has
\begin{align*}
q^{m_1+m_2-2}-q^{m_2-2}-q^{m_1}|G_{m_2}|<w_{min}<w_{max}\le q^{m_1+m_2-2}+q^{m_1}|G_{m_2}|.
\end{align*}
Especially for $m_2\ge 8$, note that $|G_{m_2}|=q^{\frac{m_2}{2}}$, we have 
\begin{align*}
\begin{aligned}
\frac{w_{min}}{w_{max}}> 1-\frac{q^{m_2-2}+2q^{m_1+\frac{m_2}{2}}}{q^{m_1+m_2-2}+q^{m_1+\frac{m_2}{2}}}> 1-\frac{q^{m_1+m_2-3}}{q^{m_1+m_2-2}+q^{m_1+\frac{m_2}{2}}}>1-\frac{1}{q+q^{3-\frac{m_2}{2}}}=\frac{q-1}{q}.
\end{aligned}
\end{align*}
Hence,  by Lemma \ref{51}, $\mathcal{C}_{D}$ is  minimal. Thus, its dual code can be employed to construct secret sharing schemes with interesting access structures.  

\subsection{Strongly regular graphs with new parameters}
Some notations and results for strongly regular graphs are given as follows \cite{RC1986}.

A connected graph of $N$ vertices is strongly regular with parameters $(N,K,\lambda,\mu)$, if it is regular with valency
$K$, and according as the two given vertices are adjacent or non-adjacent, the number of vertices joined to two given vertices is $\lambda$ or $\mu$, respectively.

Let $G=[\mathbf{y}_1,\mathbf{y}_2,\ldots,\mathbf{y}_n]$ be the generator matrix of an $[n,k]$ linear code $\mathcal{C}$ over $\mathbb{F}_q$, where $\mathbf{y}_i\in\mathbb{F}_q^k$ $(i=1,2,\ldots,n)$. Let $\mathbf{V}= \mathbb{F}_q^k$, $\mathbf{O}= \{\langle \mathbf{y}_i\rangle~|~ i = 1,2,\ldots,n\}$ and $\Omega= \{v\in \mathbf{V}~|~ \langle v\rangle\in\mathbf{O}\}$. Define a graph $G(\Omega)$ with vertices based on the vectors in $\mathbf{V}$, and
any two vertices are joint if and only if their difference is in $\Omega$. By Theorem $3.2$  \cite{RC1986}, $G(\Omega)$ is strongly regular
if and only if $\mathcal{C}$ is a projective two-weight code. Assume that the nonzero weights of $\mathcal{C}$ are just $w_1$ and $w_2$. By Corollary $3.7$  \cite{RC1986}, the parameters of $G(\Omega)$ are given in the following,
\begin{align*}
&N=q^k,\\
& K=(q-1)n,\\
&\lambda=K^2+3K-q(w_1 +w_2)-Kq(w_1+w_2)+q^2w_1w_2,\\
&\mu= K^2+K-Kq(w_1+w_2)+q^2w_1w_2.
\end{align*}

Now we give the strongly regular graphs from $\mathcal{C}_{S\times D_1}$ and $\mathcal{C}_{\tilde{D}_1}$, respectively.

By taking $m_1=m_2=m$ in Theorem \ref{t1},  $\mathcal{C}_{S\times D_1}$ is a $\big{[}\frac{(q^{m}-1)^2}{q-1}  ,2m,q^{2m-1}-2q^{m-1}\big{]}$ projective two-weight linear code with weight enumerator
\begin{align*}
1+2(q^{m}-1)z^{q^{2m-1}-q^{m-1}}+(q^{m}-1)^2z^{q^{2m-1}-2q^{m-1}}.
\end{align*}
Thus $\mathcal{C}_{S\times D_1}$ yields a strongly regular graph $G(\Omega)$ with the following parameters,{
\begin{align*}
N=q^{2m},~~ K=(q^{m}-1)^2,~~\lambda=(q^m-2)^2,~~\mu=(q^m-1)(q^m-2).
\end{align*}}
Similarly, $\mathcal{C}_{S\times\tilde{D}_1}$ can yield a strongly regular graph with the following parameters,
\begin{align*}\tiny
&N=q^{m_1+m_2},~~K=q^{m_1+m_2}-q^{m_2},~~\lambda=q^{m_1+m_2}-2q^{m_2},~~\mu=q^{m_1+m_2}-q^{m_2}.
\end{align*}

Compared with the known strongly regular graphs \cite{ZH2021,RC1986,ZH20161,CD2007}, the above strongly regular graphs are new classes.      

\section{Conclusions}
	 In this paper, several classes of projective $t$-weight $(t\in\{2,\ldots,7\})$ linear codes over  $\mathbb{F}_q$ are constructed from  defining sets, and their  weight distributions are determined by using character sums. Especially, some of these codes are Griesmer codes or distance-optimal near Griesmer codes. Furthermore, two new classes of strongly regular graphs are given from  these projective two-weight linear codes,  and some of these codes are suitable for applications in secret sharing schemes.
	 			
       \end{document}